\documentclass[aps,prl,twocolumn,groupaddress,floatfix]{revtex4-1}

\usepackage{amsmath,amssymb,graphicx}
\usepackage{color}
\usepackage{xcolor}
\usepackage[T1]{fontenc}
\usepackage[bitstream-charter]{mathdesign}

\usepackage{braket}
\definecolor{darkblue}{rgb}{0, 0, 0.8}
\usepackage[colorlinks=true, breaklinks=true, linkcolor=darkblue, citecolor=darkblue, urlcolor=darkblue]{hyperref}
\newcommand{\doilink}[2]{\href{http://dx.doi.org/#1}{#2}}
\newcommand{\urllink}[2]{\href{https://github.com/thermal-vapours/#1}{#2}}

\newcommand{\beq}{\begin{equation}}
\newcommand{\eeq}{\end{equation}}

\begin{document}

\title{Measurement of the atom-surface van der Waals interaction\\
by transmission spectroscopy in a wedged nano-cell}

\author{T. Peyrot}
\author{N. \v{S}ibali\'c}
\author{Y.R.P. Sortais}
\author{A. Browaeys}
\affiliation{Laboratoire Charles Fabry, Institut d'Optique Graduate School, CNRS,
Universit\'e Paris-Saclay, F-91127 Palaiseau Cedex, France}
\author{A. Sargsyan}
\author{D. Sarkisyan}
\affiliation{Institute for Physical Research, National Academy of Sciences - Ashtarak 2, 0203, Armenia}

\author{I.G. Hughes}
\author{C.S. Adams}
\affiliation{Department of Physics, Rochester Building, Durham University, South Road, Durham DH1 3LE, United Kingdom}

\date{\today}

\begin{abstract}
We demonstrate a method for measuring atom-surface interactions
using  transmission spectroscopy of thermal vapors confined in a wedged nano-cell.
The wedged shape of the cell allows complementary measurements of both the bulk atomic vapor and atoms close to surfaces experiencing strong
van der Waals atom-surface interaction.
These are used to tightly constrain the dipole-dipole collisional parameters of a theoretical model for transmission spectra that
accounts for atom-surface interactions, cavity effects, collisions with the surface of the cell and atomic motion.
We illustrate this method on a cesium vapor in a sapphire cell, find $C_3=1.3\pm0.1$\,kHz.$\mu$m$^3$ and demonstrate that even the weakest of the van der Waals atom-surface interaction
coefficients --- for ground-state alkali atom transitions --- can be determined with a very good precision.
This result paves the way towards a precise quantitative characterization of 
atom-surface interactions in a wide range of atom-based  nano-devices.
\end{abstract}

\maketitle

Atoms close to surfaces offer new possibilities for
engineered atom-atom and atom-light interactions through
light confinement and
surface mode excitations \cite{Chang2018,Lodahl2017,Stehle2013, Mitsch2014},
ultimately  down to the single-photon level~\cite{Chang2008,Aoki2010a,Chang2014}. 
This has stimulated a recent growth in the number of platforms where atoms are kept
close to surfaces, ranging from 
nano-fibers~\cite{LeKien2004,Vetsch2010,Deasy2014,Mitsch2014,Patterson2018} and
nano-cells~\cite{Sarkisyan2001, Peyrot2019b} to waveguides~\cite{Ritter2018} and microtoroidal 
optical resonators~\cite{Aoki2010a}.
Simultaneously, shrinking the dimensions of atom-based sensors \cite{Knappe2005,Wakai2012,Wade2016a} 
increases the number of atoms
close to a surface relative to atoms in the bulk.
Atom-surface interactions are therefore becoming increasingly
important: they may limit the ultimate achievable
precision of atom vapor sensors and they are crucial 
in understanding the dynamics in each new 
platform~\cite{Aoki2010a,Deasy2014,Patterson2018}.
However, despite their significance, direct measurements of atom-surface 
interactions are scarce.

Measuring the van der Waals (vdW) atom-surface interaction ---
that scales with the distance $z$ to the surface as $1/z^3$~\cite{Hinds1991} --- is
challenging as it requires placing the atoms in a given internal state at a distance
$z<\lambda/(2\pi)$ from the surface~\footnote{At larger distances 
interaction potential retardation effects have to be taken into account, 
resulting in Casimir-Polder potential $\propto z^{-4}$. See e.g. the review~\cite{Hinds1991}}. Here $\lambda$ is the wavelength of the strongest atomic 
transition from the considered state. 
Previous experiments on
vdW atom-surface interactions used sophisticated techniques
like reflections of
cold atoms on a surface~\cite{Landragin1996,Mohapatra2006a},
high-lying atomic states~\cite{Fichet2007} or both~\cite{Hinds1992}.
High-lying states allow easier access to the vdW regime as:
(i) transitions among higher-lying states correspond to
longer wavelengths $\lambda$, relaxing the
constrain on the atom-surface distance; and 
(ii) these transitions have 
larger dipole matrix elements, resulting in a
stronger vdW coefficient $C_3$ in the atom-surface
potential $V(z) = -C_3/z^3$. 
However, for many applications~\cite{Lodahl2017,Stehle2013, Mitsch2014,Chang2008,Aoki2010a,Deasy2014,Mitsch2014,Patterson2018,
Sarkisyan2001, Peyrot2019b,Ritter2018},
it is the properties of the ground-state atom-surface potential
that are of most interest.
 
Spectroscopy of thermal vapors contained in cells~\cite{Oria1991,Failache1999,Fichet2007,Sargsyan2017} 
is an attractive method for the measurement of atom-surface interactions,
since it can be used for a large range of vapors, atomic or molecular,
and surfaces. However, measurements have low precision for weak vdW interaction
strengths of ground state atoms, mainly limited by the uncertainty in estimating
collisional processes in dense vapors~\cite{Oria1991,Fichet2007}.
A recent method that measured the ground state
vdW interaction based on {\it fluorescence} spectroscopy in low-density 
thermal vapors~\cite{Whittaker2014} raised debate~\cite{Bloch2015}
about the absolute achievable precision of the measurements,
since the theoretical model used~\cite{Whittaker2014} 
neglected atomic motion in the spatially varying atom-surface potential. 
Finding simple and precise methods that would allow
reliable extraction of the ground-state atom-surface
potential parameters, and characterization of the atom dynamics
in the proximity of surfaces in new platforms, remains an open goal.
 
Here, we demonstrate a new method for measuring the atom-surface
interaction. Using a wedged nano-cell we obtain {\it transmission} spectra
for a vapor thicknesses range of $L=50 -275$\,nm.
This allows access both to the region where vdW interactions have 
strong effects on the total transmission signal, and the region where atoms in the bulk,
not affected by the vdW induced shifts, dominantly contribute. 
Spectra from the thick region yield the collisional parameters for
the bulk atomic vapor, thus allowing reliable fitting of thin-region spectra using our model. 
The model includes atomic motion in a spatially varying atom-surface potential,
in addition to other surface-induced transient effects~\cite{Peyrot2019a}.
This procedure allows measurement of the atom-surface potential 
even for the ground state transitions with a good precision. 
We illustrate it by a measurement of the vdW induced shift of the Cesium 
$6S_{1/2}\rightarrow 6P_{1/2}$ transition in the presence of a Sapphire surface.
Our model significantly improves on a phenomenological
one based on fitted line shift and broadening parameters, and 
shows excellent agreement with the measured spectra.
 
\begin{figure}[!t]
\includegraphics[width=\columnwidth]{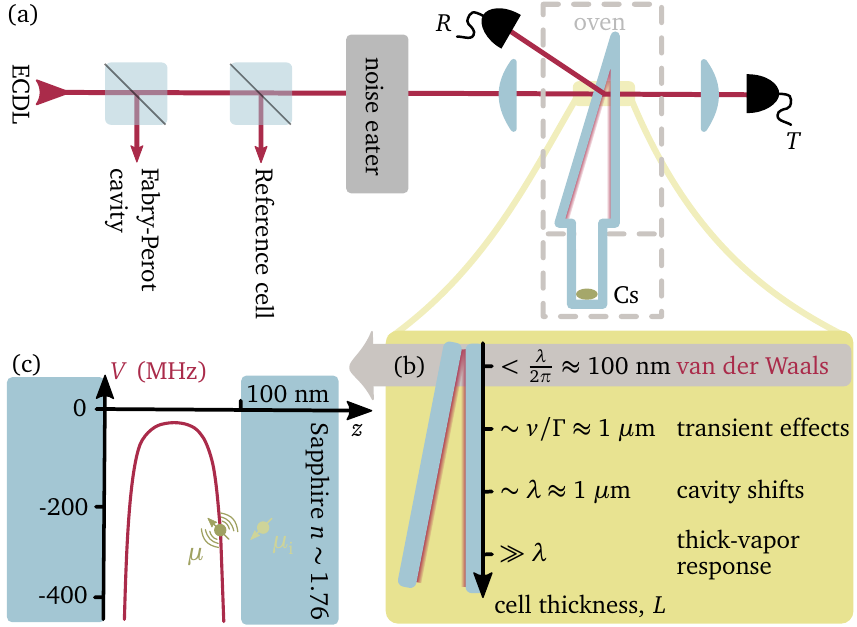}
\caption{\label{fig:outline}
(a) A frequency-calibrated, intensity-stabilized external cavity diode laser (ECDL) 
probes a cesium vapor confined in a wedged sapphire
nano-cell. The cell thickness is locally measured using interference
in the reflection signal $R$.
(b) The wedged shape of the cell allows probing the transmission $T$
in thick cell regions, where the contribution of atoms close to the 
surfaces is negligible compared to the contribution of the atoms
in the bulk. It also allows going to sub- 100~nm thick region
where the atoms are strongly
affected by the van der Waals potential, due to the interaction of their dipole
with its image in the surface (c). This gives rise to a position $z$ dependent level shift $V(z)$,
here calculated for the D1 Cs transition.}
\end{figure} 
 
The experimental setup~\cite{Peyrot2019a} is presented in Fig.~\ref{fig:outline}(a).
We use an external cavity diode laser (ECDL) to probe the $D_1$ ($\lambda = 894$\,nm) 
transition of a cesium vapor contained in an ultra-thin wedged sapphire cell, 
with thickness varying from  $L= 30$\,nm to 2\,$\mu$m. 
We measure the \emph{in-situ} vapor thickness $L$ using interference resulting from
the far off-resonant reflections from the wedge.  With a  beam waist of 50~$\mu$m ($1/\mathrm{e}^2$ intensity),
the uncertainty in the  determination of $L$ is 5~nm.
The frequency scan of the ECDL is linearized using a Fabry-Perot
cavity, and  referenced to a standard 7~cm Cesium spectroscopic cell. 
The laser intensity is stabilized during the scan by an acousto-optic modulator-based noise eater. 
The wedged nano-cell is placed in a double oven: the bottom part containing the cesium reservoir is
kept at 235~$^\circ$C, and sets the atom number density $\mathcal{N}$; 
the top part with the wedged sapphire windows is kept at a temperature 30~$^\circ$C higher to avoid
vapor condensation. The oven is temperature stabilised to $\sim$5~$^\circ$C. 
A low-noise photodiode is used to obtain the transmission signal $T$ subtracted from the background noise.
 
The theoretical analysis of transmission spectra in nano-cells is
significantly more complicated than for transmission through bulk vapor~\cite{Zentile2015}. 
With the reduction of the vapor thickness $L$, a number of effects starts to play a role [Fig.~\ref{fig:outline}(b)]: 
(i) for micrometer thick layers, the cell walls act as a low-finesse cavity, 
resulting in level shifts~\cite{Peyrot2018a}; 
(ii) the cell windows also cause dephasing of atoms upon direct collisions~\cite{Schuurmans1976}.
Atoms flying off the walls experience transient dynamics during a time $1/\Gamma$,
with $\Gamma$ the collisionally-broadened linewidth.
For a cell thickness below $v/\Gamma$ ($v$ is the average atom velocity), 
a significant number of probed atoms experience the transient regime, which significantly 
modifies the measured transmission~\cite{Peyrot2019a}. For cesium atoms at a
temperature of $\sim200~^\circ$C this corresponds to a distance  of $\sim 1~\mu$m.
Finally, (iii) at atom-surface distances $z<\lambda/(2\pi)$
atomic energy levels experience a vdW shift $V = -C_3/z^3$ [Fig.~\ref{fig:outline}(c)]. 
This atom-surface interaction comes from the coupling between an atomic
dipole and its image dipole in the sapphire surface~\cite{Fichet1995}.
To extract this vdW interaction at small atom-surface distances (iii), it is necessary to first account for
the effects (i) and (ii) that modify the atom dynamics in thin cells even
outside the range of the atom-surface potential.

We have recently developed a model for the atom response in the bulk~\cite{Peyrot2019a},
away from the influence of the vdW potential, following Refs.~\cite{Bloch2005, Dutier2003}. 
To account for the transient atoms dynamics ~(ii) described above, 
we solve the optical Bloch equations~\cite{SM_atom_surface} for the atom coherence field 
$\rho_{21}(z,v,\omega)=\rho_{+}(z,v,\omega)e^{ik z}+\rho_{-}(z,v,\omega)e^{-ikz}$, 
for each atomic velocity $v$ class, where $k=2\pi/\lambda=\omega/c$ is the laser field wave-vector {\it in vacuum}. 
Assuming a loss of coherence in atom-wall collisions ~\cite{Schuurmans1976}, one obtains
\begin{eqnarray}
\rho_{\pm}(z,v>0) &=& i\frac{d_{FF'}E_{\pm}}{2\hbar v} 
\int_{0}^{z}
\exp\left[\frac{\Lambda_\pm(z')-\Lambda_\pm(z)}{v}\right]
\mathrm{d}z' ,\label{eq:rhoV+} \\
\rho_{\pm}(z,v<0) & = & -i \frac{d_{FF'}E_{\pm}}{2\hbar v}
\int_{z}^{L}
\exp\left[\frac{\Lambda_\pm(z')-\Lambda_\pm(z)}{v}\right]
\mathrm{d}z', \label{eq:rhoV-}
\end{eqnarray}
where $E_\pm$ are the co- and counter-propagating driving fields
along the $z$ direction, and $d_{FF'}$ is the dipole moment for the
hyperfine transition $F\rightarrow F'$. In a bulk cell, away from the
range of the atom-surface potential, 
a level shift $\Delta_{\rm P}$ and a broadening $\Gamma_{\rm P}$ of 
the transition arise due to the 
atom-atom collisional interaction.
Thus, for $z>\lambda/(2\pi)$, $\Lambda_\pm$
in Eqs.~(\ref{eq:rhoV+}-\ref{eq:rhoV-}) is 
\begin{equation}\label{eq:lamdaBulk}
\Lambda^{\rm bulk}_\pm(z) =
\left[ (\Gamma_{\rm P}+\Gamma_0) /2 - i(\Delta_{FF'} \mp kv)\right]z\ ,
\end{equation}
for a laser detuning $\Delta_{FF'}= \omega-\omega_{FF'} -\Delta_{\rm P}$ from the transition at 
frequency $\omega_{FF'}$ with radiative linewidth $\Gamma_0/(2\pi)\approx 4.6$ MHz for the Cs D1 line~\cite{SM_atom_surface}. 
Equations~(\ref{eq:rhoV+}-\ref{eq:rhoV-}) can then be analytically solved,
and the polarisation of the medium $P(z,\omega) = \sum_{F,F'}\int_{-\infty}^{\infty} {\rm{d}}v~2\mathcal{N}M_{\rm b}(v)d_{FF'}\rho(z ,v, \omega)$ is 
obtained by summing the contributions from all velocity
classes in the Maxwell-Boltzman distribution $M_{\rm b}(v)$.

In order to include the cavity effects (i),
we take into account the transmission $t_1 = 2 n/(1+n)$ and reflection $r_2 = (1-n)/(1+n)$ 
coefficients of the driving field $E_0$ at the two sapphire cell walls (refractive index $n$).
Assuming an optically dilute atomic medium,  we neglect the contribution of atoms to 
the driving field $E_\pm$ \emph{inside the cavity},
so that $E_{+}e^{ikz}+E_{-}e^{-ikz} \simeq t_1/[1-r_2^2\exp(2i k L)]E_0\{\exp[i k z]+r_2\exp[i k (2L-z)]\}$.
Similarly, the radiated atomic fields $E_{\rm A+}$ and $E_{\rm A-}$ 
initially co- or counter-propagating
along the $z$-axis respectively~\cite{SM_atom_surface},
can be reflected multiple times inside the cavity~\cite{Peyrot2019a}. They give rise to an atom induced field
outside the cavity 
\begin{equation} \label{Field_dipole}
E_{\rm A\pm}(z)={\frac{t_2}{1-r_2^2e^{2i k L}}}\frac{i k }{2\varepsilon_0}\int_{0}^{L}
\mathrm{d}z' P(z',\omega)\exp[\pm i k  (z-z')]~,
\end{equation}
in the direction of initial emission,
and fields $r_2\exp[2ik(L-z)]E_{\rm A+}$ and $r_2\exp(2ikz)E_{\rm A-}$
respectively in the opposite direction~\cite{SM_atom_surface},
where $t_2 = 2/(1+n)$.
Finally, we obtain the transmission factor through the thin cell system $T=\left|E_{\rm T}/E_{0+}\right|^2$, where
$E_{\rm T}=E_{0+}+E_{\rm A+}+r_2\exp(2ikz)E_{\rm A-}$
is a superposition of the transmitted atom-induced field
and transmitted driving field $E_{0+}= t_1t_2/[1-r_2^2\exp(2i k L)]E_0 \exp(i k z)$.

\begin{figure}[!t]
\includegraphics[width=\columnwidth]{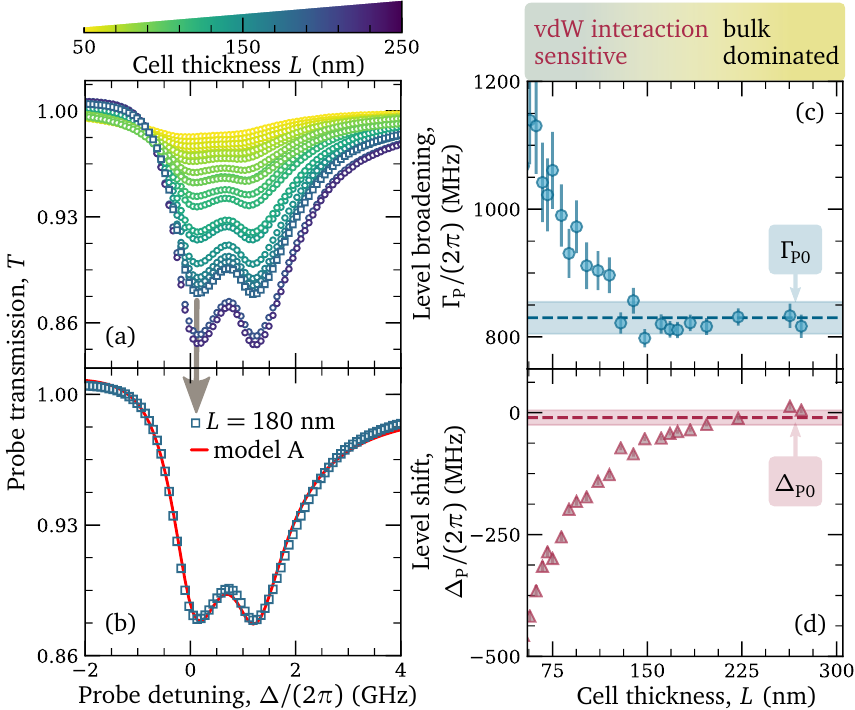}
\caption{\label{fig:spectra}
Determination of the collisional broadening $\Gamma_{\rm P0}$ and 
shift $\Delta_{\rm P0}$ in the bulk vapor.
(a) Measured transmission spectra at 235~$^\circ$C,
for cell thickness $L$ in range 50--225~nm. The left and right
peaks correspond to the $F=4\rightarrow 3$ and $F=4\rightarrow4$ 
transitions of the Cs D1 line respectively.
Model A, not including atom-surface interactions, fits well
spectra in the thick part of the cell (b), allowing the extraction of
bulk properties of the vapor from values obtained for large $L$ (c--d).
For small $L$, the influence of the atom-surface 
interactions appears as an additional thickness-dependent line 
broadening and transition shift.}
\end{figure}

The model described above (from now on Model A) fits well the measured
transmission spectra [Fig.~\ref{fig:spectra}(a)] for cell thicknesses $L\ge 150$~nm~[see for example Fig.~\ref{fig:spectra}(b)],
where the signal is dominated by atoms far from the surface. 
In this region, fitting the temperature, level shift $\Delta_{\rm P}$ and broadening
$\Gamma_{\rm P}$ allows us to obtain the collisional
self-broadening $\Gamma_{\rm P0}/(2\pi)=830 \pm 10$~MHz [Fig.~\ref{fig:spectra}(c)] and line shift 
$\Delta_{\rm P0}/(2\pi)=-10 \pm20$~MHz 
[Fig.~\ref{fig:spectra}(d)] \emph{in the bulk vapor}, 
arising solely from atom-atom collisions. The value and error bar are the mean and standard error of the fitted values for cell thicknesses $L\ge 175$ nm. The measured broadening is
in good agreement with theoretical predictions~\cite{Weller2011} at $235~^{\circ}$C.
  
\begin{figure}
\includegraphics[width=\columnwidth]{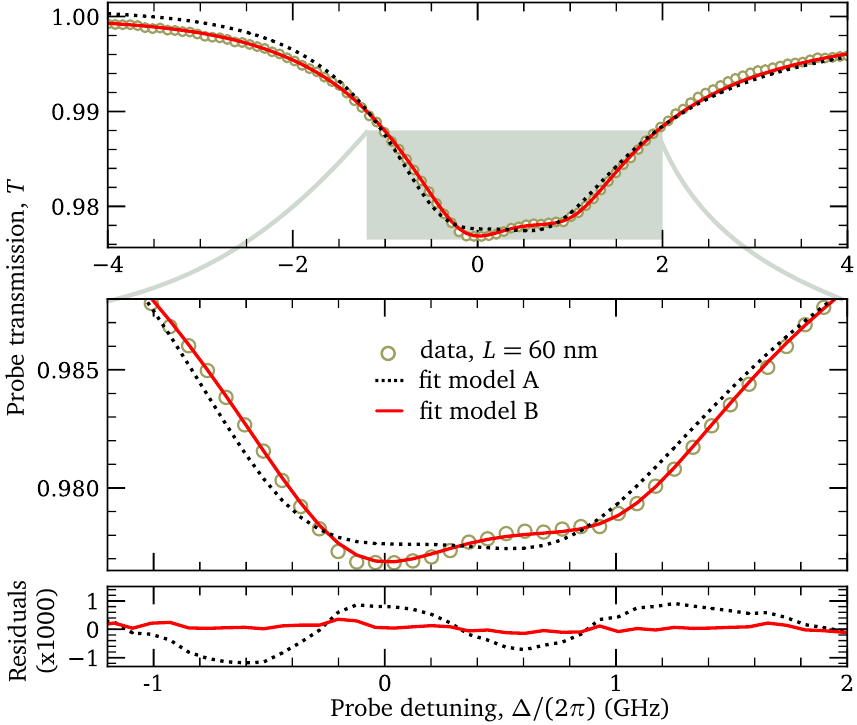}
\caption{\label{fig:modelAvsB} 
Comparison of Model B that accounts for vdW interactions with phenomenological Model A.
Model A (dotted line), not including atom-surface interactions,
misses line-shape features 
for spectra in the thin cell region (circles), as highlighted on the zoomed inset.
This is despite the fact that in addition to temperature, we take the
broadening $\Gamma_{\textrm{P}}$ and shift $\Delta_{\textrm{P}}$ 
as free parameters to 
phenomenologically account for atom-surface interactions. 
Model B (solid line) --- that includes atom motion in the spatially dependent
van der Waals potential explicitly ---
reproduces the asymmetric double-peak feature 
perfectly, with only the temperature and $C_3$ as free parameters when imposing 
bulk-determined line shift $\Delta_{\rm P}=\Delta_{\rm P0}$ and broadening
$\Gamma_{\rm P}=\Gamma_{\rm P0}$. }
\end{figure}

For cell thicknesses $L\lesssim 150$~nm, the contribution from atoms close to the
surface becomes significant. 
The vdW interaction~[Fig.~\ref{fig:outline}(c)] offsets the $6S_{1/2}$ and $ 6P_{1/2}$ levels 
by different amounts due to the different vdW $C_3$ coefficients  for the two states.
Model A captures this atom-surface interaction \emph{phenomenologically} 
as a cell thickness dependent shift and broadening, as shown in Fig.~\ref{fig:spectra}(c-d). 
However, this phenomenological fit does not provide direct access to the
$C_3$ vdW coefficient. 
In addition, Fig.~\ref{fig:modelAvsB} indicates that fine features
of thin-cell spectra are not captured by Model A. 
This motivates extending the model to explicitly include the spatially varying
vdW potential.

To do so,
we now make the transition frequency spatially dependent:
$\omega_{FF'} \rightarrow \omega_{FF'} -V(z)$. 
The atom-surface position dependent shift of the D1 transition
frequency is $V(z)=-C_3[1/z^3+1/(L-z)^3]$, where $C_3 =  C_3[6P_{1/2}]-C_3[6S_{1/2}]$. 
In summing independently the potentials of each surface, 
we account only for the interaction of the atomic dipole
with its image from the two surfaces. 
Image dipoles of image dipoles
have contributions smaller than $\sim 3$\%~\cite{SM_atom_surface}. 
Solving the Bloch equations leads to the 
solution for the atomic coherence field of the same form as in
Eqs.~(\ref{eq:rhoV+}-\ref{eq:rhoV-}), except that now
\begin{eqnarray}\label{eq:lambdaFull}
\Lambda_{\pm}(z) &=  & \left\{ \frac{\Gamma_0+\Gamma_{\rm P0}}{2}
-i (\omega-\omega_{FF'}-\Delta_{\rm P0}\mp kv) \right. \nonumber \\
& & ~\left. +i C_3\left[\frac{1}{2z^3}-\frac{1}{2z(L-z)^2}\right] \right\}z.
\end{eqnarray}
Here we set the collisional broadening $\Gamma_{\rm P0}$
to be the same as in the bulk,
which is justified as we are far from polaritonic resonances of the crystal surface~\cite{Failache2002}. 
The first part of~Eq.~(\ref{eq:lambdaFull}) describes the bulk~[Eq.~(\ref{eq:lamdaBulk})],
while the second line is due to the atom-surface vdW potential.
By using Eq~(\ref{eq:lambdaFull})
in Eq.~(\ref{eq:rhoV+}-\ref{eq:rhoV-}), we include the
\emph{internal} atomic dynamics
as the atoms move relative to the surface, thus experiencing
\emph{time-varying level shift} due to the vdW interaction $V(z)$,
to obtain the coherence field for each velocity class at location $z$. In previous works
based on fluorescence measurement~\cite{Whittaker2014}, 
the motion was not accounted for~\cite{Bloch2015}.
For details of the numerical integration see Supplemental~\cite{SM_atom_surface}. We call this Model B.

\begin{figure}
\includegraphics[width=\columnwidth]{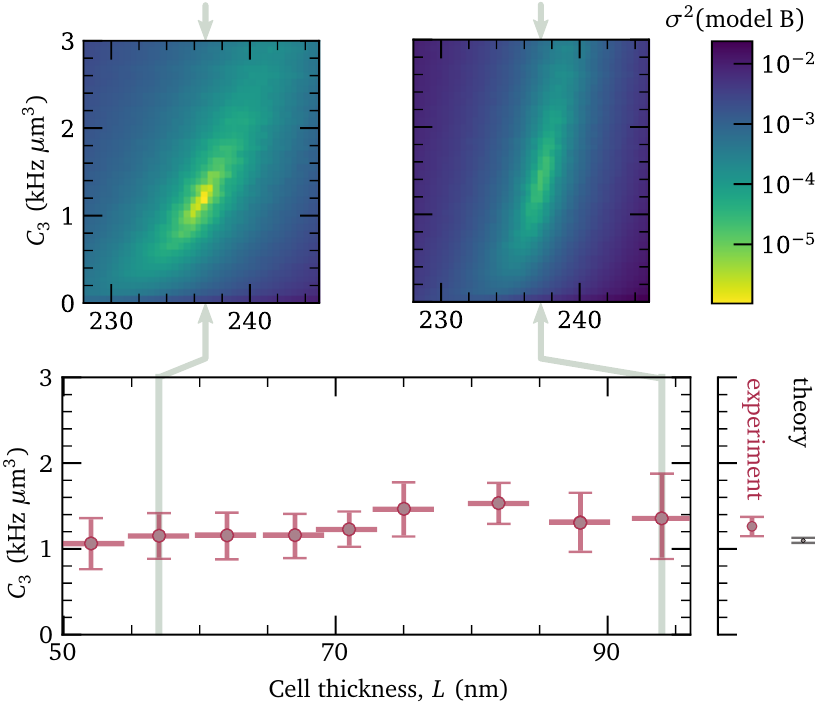}
\caption{\label{fig:C3determination} 
Determination of the $C_3$ coefficient for the D1 line of Cs atoms close to the sapphire surface.
Main plot: $C_3$ (circles) obtained by fitting Model B to the transmission spectra for
different cell thickness. Error bars are systematic.
Top insets: temperature-$C_3$ maps of the sum of squared residuals $\sigma^2$
for Model B, for two cell thicknesses indicated by vertical shaded bars on the main plot.
The final value of $C_3$, obtained as a weighted average,  is in excellent
agreement with the theoretical prediction (side scale).}
\end{figure}

The result of the fitting of the data for a thin cell with $L=60$~nm is shown in Fig.~\ref{fig:modelAvsB}. 
Model B has only two free parameters, the temperature and $C_3$, 
as $\Gamma_{\rm P0}$ and $\Delta_{\rm P0}$ are constrained to their bulk extracted values. 
Yet, it shows outstanding agreement compared to Model A that only phenomenologically
accounts for atom-surface vdW interactions. In particular, the red-side asymmetry 
cannot be retrieved without taking properly the vdW shift into account. 
The value for $C_3$ is extracted from fitting the spectra for each
cell thickness $L$ (Fig.~\ref{fig:C3determination} main panel).
Thin cells yield a larger fraction of atoms close to the
surface, where the vdW potential induces large level shifts, and allow tighter
constraints on the fitted parameters. This is visible on the map of the sum
of squared residuals (Fig.~\ref{fig:C3determination} top insets),
where the region that minimizes $\sigma^2$ is more
tightly localized along the $C_3$-axis for thin cells. The error bars on $C_3$ (main panel of Fig.~\ref{fig:C3determination}) are systematic and result from propagating errors on the cell thickness and the fitted bulk vapor parameters $\Gamma_{\textrm{P0}}$ and $\Delta_{\textrm{P0}}$~\cite{SM_atom_surface}. Assuming no dependence on $L$, the final value for $C_3$ is obtained as an average of the fitted values
for different cell thicknesses weighted by their individual error 
bars~\cite{Hughes2010}.
We obtain $C_3=1.3\pm0.1$\,kHz.$\mu$m$^3$, in good agreement~(Fig.~\ref{fig:C3determination} side scale) 
with the theoretical value $C_3^{\rm th} = 1.10\pm 0.03$\,kHz.$\mu$m$^3$ predicted from 
the Lifshitz theory of vdW interactions~\cite{Lifshitz1956,SM_atom_surface},
based on the refractive index of sapphire~\cite{Weber2003} and the latest calculations of
dipole matrix elements for low-lying cesium transitions~\cite{Safronova2016,Sibalic2016d}.

In conclusion, we have demonstrated a method to measure the interaction between an atom and
a surface. It relies on comparing the 
transmission of a laser beam through a wedged nano-cell filled with an atomic vapor to a theoretical model 
including the atom-surface interaction, cavity effects, collisions with surfaces of the cell and atomic motion. 
The wedged cell provides simultaneous access to the bulk vapor properties, 
tightly constraining theoretical parameters, and to thin cell regions with strong vdW interaction. 
We have illustrated the method with thermal cesium atoms confined within a sapphire cell
and measured the D1 transition shift due to the cesium-sapphire interaction.
Our model yields transmission  predictions in excellent agreement with the experimental data. 
The measured  $C_3$ coefficient is consistent with the non-resonant
Lifshitz theory.
Further improvements in precision are expected from adapting this work
to fluorescence measurements, as better signal to noise ratio lead to a better determination of collisional properties as well as a reduction of statistical errors in extremely thin regions.
Finally, this work opens the way to quantitative analysis of atom dynamics close to surfaces in new platforms~\cite{Aoki2010a,Deasy2014,Patterson2018},
the search for predicted atom-surface repulsive potential~\cite{Failache1999} and bound-states~\cite{Lima2000},   
and the examination of short-~\cite{Vargas1998} and long-range~\cite{Laliotis2018}
limits of atom-surface potential, using thermal vapor spectroscopy. The data and the code for theoretical model are both available \cite{PaperData,PaperCode}.

\begin{acknowledgments}
We thank A.~Laliotis and J.~Keaveney for fruitful discussions. 
T.~Peyrot is supported by the DGA-DSTL fellowship 2015600028. 
N.~\v{S}. is supported by EU Horizon 2020 (COQUDDE, Marie Sk{\l}odowska-Curie fellowship No 786702).
We also acknowledge financial support from EPSRC (grant EP/R002061/1) and Durham University. 
\end{acknowledgments}


\begin{thebibliography}{45}
\bibitem{PaperData}
Data is available at  
\doilink{10.5281/zenodo.2653099}{10.5281/zenodo.2653099}
\end{thebibliography}

%

\end{document}


\title{Supplemental Material: Measurement of the atom-surface van der Waals interaction\\
by transmission spectroscopy in wedged nano-cell}

\author{T. Peyrot}
\author{N. \v{S}ibali\'c}
\author{Y.R.P. Sortais}
\author{A. Browaeys}
\affiliation{Laboratoire Charles Fabry, Institut d'Optique Graduate School, CNRS,
Universit\'e Paris-Saclay, F-91127 Palaiseau Cedex, France}
\author{A. Sargsyan}
\author{D. Sarkisyan}
\affiliation{Institute for Physical Research, National Academy of Sciences - Ashtarak 2, 0203, Armenia}

\author{I.G. Hughes}
\author{C.S. Adams}
\affiliation{Department of Physics, Rochester Building, Durham University, South Road, Durham DH1 3LE, United Kingdom}

\date{\today}

\date{\today}

\maketitle

Here we detail the derivation of the theoretical model introduced in the main text, before discussing
the error analysis and the theoretical predictions for the van der Waals (vdW) $C_3$ coefficient. 
In Sec.~\ref{Sec:NLmodelA}, we derive the atom coherence field
inside the nano-cell, starting from the optical Bloch equations and 
including the vdW atom-surface interaction, the atomic motion,
and the atom-surface collisions. Based on that result, we calculate in 
Sec.~\ref{Sec:NLmodelB} the optical field transmitted through a thin layer of atomic vapor,
 including optical cavity effects induced by the walls of the nano-cell.
We present in Sec.~\ref{Sec:ErrorBar} the procedure for estimating the error bars 
on the experimentally measured $C_3$.
Finally the theoretical calculation of $C_3^{\rm th}$, and the numerical
values of the dipole matrix elements used are presented in Sec.~\ref{Sec:Theory}. 

\section{Derivation of the vapor-cell transmission}\label{Sec:NLmodel}

In this Section, we derive the expression of the transmission through the atomic vapor 
slab, including the optical cavity effect
due to the sapphire windows of the nano-cell and the vdW atom-surface interaction. 
We start by deriving the expression of the polarisation $P (z,\omega)$ 
of the ensemble of atoms and we subsequently compute the transmission through the nano-cell. 
The derivation is done in the limit of an optically thin vapor.

\subsection{Calculation of the vapor polarisation $P$}\label{Sec:NLmodelA}
 
We use a 1D model for the propagation along the $z$ axis of the  field through the vapor. 
We decompose the linearly polarized, monochromatic field driving the atom inside the 
cavity formed by the sapphire plates
\begin{equation}\label{eq:driving_field}
E(z,t)=\frac{1}{2}\left[E(z)e^{-i\omega t}+E^*(z)e^{i\omega t}\right],
\end{equation}
and the polarisation of the atomic vapor
\begin{equation}\label{eq:poldecomposition}
P(z,t)=\frac{1}{2}\left[P(z)e^{-i\omega t}+P^*(z)e^{i\omega t}\right],
\end{equation}
in their positive and negative frequency components.
We first consider a given atomic transition
between ground and excited states, with respective 
hyperfine states $F$ and $F'$. The transition frequency is
$\omega_{FF'}$, the total homogeneous linewidth $\Gamma_{\rm t}$,
and the dipole matrix element $d_{FF'}$ for the driven transition
by the linearly polarizated laser field.
For this transition, we introduce the classical coherence field
$\rho_{21}(z,t,v) = \langle \rho_{21}^{(i)}(t,v) \rangle_{\textrm{atoms }} $ where $\langle \ldots \rangle$
is the configuration average of the coherences $\rho_{21}^{(i)}$ of all atoms $i$ located within the $(z, z+\delta z)$ slab 
and with velocities in the range $(v, v+\delta v)$ where
$\delta z, \delta v \rightarrow 0$.
The evolution of this field is given by the hydrodynamic equation:

\beq \label{Maxwell-Bloch2}
\frac{\partial \rho_{21}(z,t,v)}{\partial t} =
\underbrace{
-i\omega_{FF'} \rho_{21}(z,t,v)+i\frac{d_{FF'} E(z,t)}{\hbar}-\frac{\Gamma_{\rm t}}{2} 
\rho_{21}(z,t,v)}_{\textrm{Internal atom evolution (Bloch equations)}} 
\underbrace{- v\frac{\partial \rho_{21}(z,t,v)}{\partial z}}_{\textrm{atoms flying into and out from the slab at position }z}~ .
\eeq

\noindent The first part of the right hand side of the equation is the
standard Bloch equations for the evolution of the coherence under
weak driving, when the population
of the excited state can be neglected ($\rho_{22}\rightarrow 0 $)~\cite{Aspect2010}.
In addition, since the field is defined as an average over many atoms at a given
location $z$, it can also change due to atoms flying into and out of
the vapor slab at position $z$. This is accounted for by the
second part of  the right-hand side of the Eq.~(\ref{Maxwell-Bloch2}), which appears  
owing to the fact that we consider the coherence {\it field} rather than the  coherence
of a given atom. It is valid assuming that the vapor is homogeneous,
i.e. the atom number density $\mathcal{N}$ for a given velocity 
is the same everywhere in the vapor [$\partial \mathcal{N}(z,v)/\partial z=0$].
This is a very good approximation as long as
the kinetic energy of the atoms is much larger than the potential
energy induced by the vdW interactions,
i.e. for $z>(2C_3/k_B T)^{1/3} = 0.6$~nm.
Using Eq.~(\ref{eq:driving_field}) and the
rotating wave approximation, we obtain the time-independent equation
\beq \label{Full_Coherence}
v\frac{\partial \rho_{21}(z,v)}{\partial z}=
-\left[\frac{\Gamma_{\rm t}}{2}-i\Delta_{FF'}\right] \rho_{21}(z,v)+i\frac{d_{FF'}E(z)}{2\hbar} \ ,
\eeq
where $\Delta_{FF'} = \omega-\omega_{FF'}$ is the laser
frequency detuning from the atomic transition. For the general driving field
\beq \label{Fieldplusmoins}
E(z)=E_{+}e^{ikz}+E_{-}e^{-ikz},
\eeq
consisting of the co-propagating $E_+$ and counter-propagating
$E_-$ field along the $z$ axis with $k=\omega /c $ the laser wave-vector, we write the coherence as
\beq \label{Coherenceslusmoins}
\rho_{21}(z,v)=\rho_{+}(z,v)e^{ikz}+\rho_{-}(z,v)e^{-ikz}.
\eeq
We thus obtain the following two equations from Eq.~\eqref{Full_Coherence} 
\begin{eqnarray}
v\frac{\partial \rho_{+}}{\partial z} & = & -\left[ \Gamma_{\rm t} /2-i(\Delta_{FF'} -kv)\right]\rho_{+}+i\frac{d_{FF'}E_{+}}{2\hbar} , \label{Eq:Coherences+} \\
v\frac{\partial \rho_{-}}{\partial z} & = &-\left[ \Gamma_{\rm t} /2-i(\Delta_{FF'} +kv)\right]\rho_{-}+i\frac{d_{FF'}E_{-}}{2\hbar} .
\label{Eq:Coherences-}
\end{eqnarray}

In the two expressions above, both $\Gamma_{\rm t}$ and $\omega_{FF'}$
may depends on $z$, for example due to the atom-surface interaction potential. Using the
variation of constant method for differential equations, we find  for
$v\neq 0$ 
\begin{eqnarray} 
\rho_{+}(z,v) & = & \rho_{+}(z_0,v)e^{-\Lambda_{+}/v} 
+ i\frac{d_{FF'} E_{+}}{2\hbar v}
\int_{z_0}^{z} \mathrm{d}z'~\exp \left[\frac{\Lambda_+(z')-\Lambda_+(z)}{v}\right], \label{Solvposa:Coherences+} \\
\rho_{-}(z,v) & = & 
\rho_{-}(z_0,v)e^{-\Lambda_{-}/v}
+ i\frac{d_{FF'}E_{-}}{2\hbar v}
\int_{z_0}^{z} \mathrm{d}z'~ \exp \left[\frac{\Lambda_-(z')-\Lambda_-(z)}{v}\right],\label{Solvposa:Coherences-}
\end{eqnarray}
where the primitive:
\beq\label{Eq:lambda_pm_int}
\Lambda_\pm(z) = \int^{z} {\Gamma_{\rm t}(u) \over 2} - i[\Delta_{FF'}(u) \mp kv]\,  \mathrm{d}u\ .
\eeq 
Assuming a loss of coherence resulting from the collisions of the atoms with the cell walls~\cite{Schuurmans1976} 
(quenching collisions) leads to the boundary conditions  $\rho_{\pm }(v>0,z=0)= \rho_{\pm }(v<0,z=L)=0$.
Three cases can be distinguished: 
\begin{enumerate}
\item[(i)] $v>0$
\begin{eqnarray}
\rho_{+}(z,v>0) & = & i\frac{d_{FF'}E_{+}}{2\hbar v}
\int_{0}^{z} \mathrm{d}z'~
\exp\left[\frac{\Lambda_+(z')-\Lambda_+(z)}{v}\right],
 \label{Solvpos:Coherences+} \\
\rho_{-}(z,v>0) &=& i\frac{d_{FF'}E_{-}}{2\hbar v}
\int_{0}^{z} \mathrm{d}z'~
\exp\left[\frac{\Lambda_-(z')-\Lambda_-(z)}{v}\right];
\label{Solvpos:Coherences-}
\end{eqnarray}

\item[(ii)] $v<0$:
\begin{eqnarray}
\rho_{+}(z,v<0) &=& -i\frac{d_{FF'}E_{+}}{2\hbar v}
\int_{z}^{L} \mathrm{d}z'~
\exp \left[\frac{\Lambda_+(z')-\Lambda_+(z)}{v}\right] ,
\label{Solvneg:Coherences+}\\
\rho_{-}(z,v<0) &=& -i\frac{d_{FF'}E_{-}}{2\hbar v}
\int_{z}^{L} \mathrm{d}z'~
\exp \left[\frac{\Lambda_-(z')-\Lambda_-(z)}{v}\right];
 \label{Solvneg:Coherences-}\\
\end{eqnarray}

\item[(iii)] $v=0$:
\begin{equation} \label{Solvnull:Coherences+-}
\rho_{\pm}(z,v=0)= 
\frac{id_{FF'}E_{\pm}/2\hbar}{\Gamma /2- i\Delta_{FF'}}.
\end{equation}

\end{enumerate}

Now, knowing the coherences in Eq.~(\ref{Coherenceslusmoins}), we calculate the 
polarisation given by Eq.~(\ref{eq:poldecomposition}) by 
integrating the total atomic dipole moment over the Maxwell-Boltzmann velocity distribution $M_{\rm b}(v)$, 
and adding all relevant transitions $F\rightarrow F'$ :

\beq\label{Eq:Polarisation}
P(z,\omega)= \sum_{F,F'}\int_{-\infty}^{\infty} 2 \mathcal{N} M_{\rm b}(v) d_{FF'}
\left[ \rho_{-}(z,v,\omega)e^{-ikz}+\rho_{+}(z,v,\omega)e^{+ikz} \right] \mathrm{d}v.
\eeq

When the vdW potential is not taken into account, $\omega_{FF'}$ and $\Gamma_{\rm t}$ are independent of $z$. 
In this case, Eq. \eqref{Eq:Polarisation} gives analytical expressions that we derived in Ref.~\cite{Peyrot2019a}. 
Here instead, we consider the atom-surface potential. As a consequence, 
a specific optical transition for an atom located at a distance $z$ from the first surface is shifted as  
$\omega_{FF'}(z)=\omega_{FF'}-\left[C_3/z^3+C_3/(L-z^3)\right]$. 
Far from any polaritonic resonances of the crystal surface, 
the linewidth may however be considered as independent of $z$ and 
$\Gamma_{FF'}=\Gamma_0$ \cite{Failache2002}.
We can then calculate the integrals of Eq.(\ref{Eq:lambda_pm_int}):
\begin{equation}
\Lambda_{\pm}=\left\{\frac{\Gamma_{\rm t}}{2}-i\left[\Delta \mp kv -\frac{C_3}{2z^3} + 
\frac{C_3}{2z(L-z)^2}\right]\right\}z\ .
\end{equation}
The expression now diverges for $z=0$ and $z=L$. Besides none of equations \eqref{Solvpos:Coherences+},
\eqref{Solvpos:Coherences-},\eqref{Solvneg:Coherences-} and \eqref{Eq:Polarisation} provide analytical solutions.
These integrals are regularised  by introducing a minimal cutoff distance $L_{\rm cut}=0.1$\,nm 
from both surfaces. We have checked that it does not influence the results of the numerical computation. 

\subsection{Calculation of the driving and transmitted fields in the presence of the cavity}\label{Sec:NLmodelB}

\begin{figure}[!t]
\includegraphics[width=0.55\columnwidth]{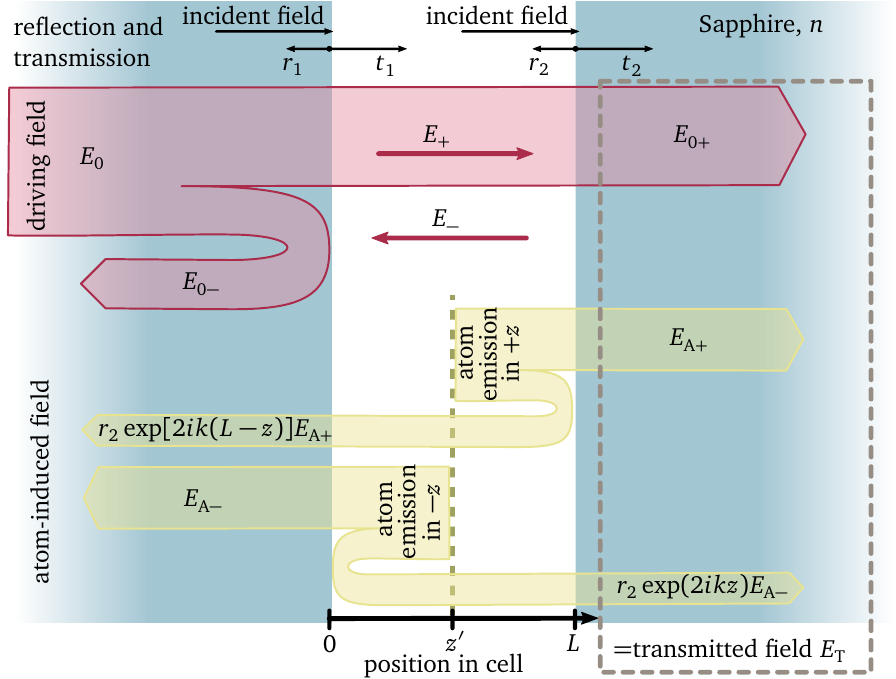}
\caption{\label{fig:all_fields}
Reflection and transmission of fields on sapphire-vapor and
vapor-sapphire interfaces. Incident driving field $E_0$ is split by this
interface into a transmitted field $E_{0+}$ and a
reflected field $E_{0-}$. Inside the low-finesse cavity formed by these
two interfaces, the driving field gives rise to a field $E_+$ propagating
along the $z$-axis direction, and a counter-propagating field $E_-$.
The field emitted by the atoms \emph{inside the cavity}
in the forward direction after multiple reflections by the cavity results in the fields
$E_{\rm A +}$ and $r_2\exp[2ik(L-z)]E_{\rm A +}$ propagating
\emph{outside the cavity} in the $+z$ and $-z$ directions respectively.
Similarly, the field  emitted initially by the atoms in $-z$ direction
results in the fields $E_{\rm A-}$ and $r_2\exp(2ikz)E_{\rm A-}$
propagating \emph{outside the cavity} in the $-z$ and $+z$ directions respectively.}
\end{figure}

The cell walls present two interfaces for the incoming driving
light $E_0$ (Fig.~\ref{fig:all_fields}). First from the sapphire into the
vacuum filled with the atomic vapor, with amplitude transmission
$t_1=2n/(1+n)$ and reflection $r_1=(n-1)/(1+n)$ coefficients given by the sapphire
refractive index $n$. This is followed by the vapor-sapphire interface,
with respective amplitude transmission $t_2=2/(1+n)$ and
reflection $r_2=(1-n)/(1+n)$ coefficients. In fact, multiple reflections of the
driving field between the two nearly-parallel sapphire surfaces that form 
a low finesse cavity result in a total driving field \emph{inside the cavity}, co-propagating along the $z$-axis direction:
\begin{equation}
E_+=E_0\frac{t_1}{1-r_2^2e^{2ikL}}\ ,
\end{equation}
and a counter-propagating driving field
\begin{equation}
E_-=E_0\frac{t_1r_2e^{2ikL}}{1-r_2^2e^{2ikL}}\ .
\end{equation}
Here $k = \omega/c$ is the field wave-vector {\it in vacuum}, as we assume that the vapor is 
optically thin so as to neglect its index of refraction. The  driving field {\it after} the cavity is therefore
\begin{equation}
E_{0+}=E_0\frac{t_1t_2}{1-r_2^2e^{2ikL}}\ . 
\end{equation} 

We now account for the field scattered by the polarisation of the
atomic vapor. The field emitted by the atoms \emph{inside the cavity}
along the $z$ axis ($+z$ direction), experiences multiple reflections 
between the two sapphire windows (see Fig.~\ref{fig:all_fields}),
resulting in the fields  
$E_{\rm A+}$ and $r_2\exp[2ik(L-z)]E_{\rm A +}$ propagating
\emph{outside the cavity} in the 
$+z$ and $-z$ direction respectively, where
\begin{equation}\label{eq:Ea+}
E_{\rm A+}= \frac{t_2}{1-r_2^2e^{2 i k L}} {ik\over 2\varepsilon_0}\int_0^L dz'~ P(z',\omega) ~e^{ik(z-z')}.
\end{equation}
Similarly the field
originally emitted by the atoms \emph{inside the cavity}
in $-z$ direction, after multiple reflections by the cell windows
results in fields $E_{\rm A-}$ and $r_2\exp(2ikz)E_{\rm A-}$ propagating
\emph{outside the cavity} in $-z$ and $+z$ directions, where
\begin{equation}\label{eq:Ea-}
 E_{\rm A-}=\frac{t_2}{1-r_2^2e^{2 i k L}} {ik\over 2\varepsilon_0}\int_0^L dz'~ P(z',\omega)~ e^{-ik(z-z')}.
\end{equation}

The transmitted field along the $z$-axis is then given by the sum of
the driving field and atom-induced field contributions
\begin{equation}
E_{\rm T+} = E_{0+} + E_{\rm A+} + r_2\exp(2ikz) E_{\rm A-}.
\end{equation}
The relative transmission is thus $T=\left| E_{\rm T}/E_{0+}\right|^2$.

\textsc{Note.} We highlight one nano-cell peculiarity in
comparison to more standard, thick cells,
regarding directionality of emitted light.
Using expression for polarisation Eq.~(\ref{Eq:Polarisation}) in
Eq.~(\ref{eq:Ea+}) for emission along the laser beam, we obtain
\begin{equation}
E_{\rm A +} \propto \int_0^L \rho_+ \mathrm{d}z' + \int_0^L \rho_- \mathrm{e}^{-2ikz'}\mathrm{d}z'.
\end{equation}
For thick cells ($L\gg \lambda$) and slowly varying coherence fields
$\rho_+$ and $\rho_-$, we note that rapid integrand sign change
in the second term prevents contribution of $\rho_-$ to $E_{\rm A+}$.
We obtain that emission in the $z+$ direction is then 
dominated by the $\rho_+$ part of the coherence field.
However for nano-cells $L\ll \lambda$, 
exponential factor is practically constant over the length of the cell,
and we see that 
\emph{$\rho_-$ also contributes to the emission in the $z$+ direction}.
This is expected if one keeps in mind that large atomic ensambles
can emit light in a given direction thanks
to the interference [given by Eq.~(\ref{eq:Ea+})]
from phase grating (c.f. Bragg grating)
imprinted by the driving laser beam into the spatial variation of coherence phase.
However, cells $L\ll \lambda$ are not thick enough to contain
a single period $\lambda$ of this phase variation (Bragg grating),
and emission integration in Eq.~(\ref{eq:Ea+}) will not single out 
contribution of $\rho_+$ to polarisation [Eq.~(\ref{Eq:Polarisation})].
Instead $\rho_-$ will contribute to emission in forward direction too.
In particular, note that this
is not the consequence of reflection of the light emitted along $z-$
on the interface. Indeed the effect would exist even in the absence of
cavity induced reflections.
This effect can also be seen as a consequence of constrain
on space-bandwidth product of Fourier transform in space,
that for small total distances
cannot distinguish between $+k$ and $-k$ components
(c.f. Heisenberg relations), giving reduction of phase  matching 
constrain for thin cells.

\section{Estimation of the error bars on the extracted $C_3$ value}\label{Sec:ErrorBar}

In this section, we explain how we have assigned the error bars for the $C_3$ coefficients 
extracted from the  transmission spectra corresponding to various thicknesses presented in Fig.(4) of the main text. 
We subsequently derive the final value and uncertainty for the atom-surface coefficient.  

To extract the transmission in intensity from the raw data, we need 
to divide the measured transmitted intensity by the intensity far from the atomic resonance.
In normalizing the data, an issue arises due to a residual variation of the intensity not corrected 
by the stabilisation of the laser power during the scan across the atomic resonances. This residual 
variation may distort the lineshape and introduce unphysical shifts.
Our normalization procedures therefore accounts for this nearly linear intensity variation. 
We have checked that a small variation of the residual slope of the intensity variation 
does not influence significantly the fitted $C_3$ value (on the order of few percent). 
We then checked that the statistical repetition of the measurements does not modify the extracted $C_3$ value.

As explained in the main text, in order to extra the $C_3$ coefficient, we fix the thickness measured with an uncertainty of 5 nm. 
Collisional  linewidth and shift are also fixed to their bulk values  $\Delta_{\rm{P}0}$ and $\Gamma_{\rm{P0}}$
measured at large cell thicknesses. This values are the average of the shifts $\Delta_{\rm{P}}$ 
and broadenings $\Gamma_{\rm{P}}$ obtained 
for thicknesses  $L\ge175$nm, and the error bar on these parameters is the standard deviation of this ensemble of
values. 
In order to assign an error bar to the $C_3$ coefficient, we 
perform an error propagation in the following way:
we vary within their error intervals $\Delta_{\rm{P}}$ and $\Gamma_{\rm{P}}$ and $L$
and repeat the fitting procedure to find the $C_3$
that minimizes the sum of squared residuals.
We thus obtain the systematic error on $C_3$ associated to the uncertainty in the determination
of $\Delta_{\rm{P}0}$ and $\Gamma_{\rm{P0}}$. 
For each cell thickness $L$ we then  assign an error bar that
encompasses all the  $C_3$ values obtained in this way.

When applying this procedure, we find that the extracted $C_3$ is insensitive 
to the value of $\Gamma_{\rm {P0}}$. On the contrary, it is very sensitive to the value of $\Delta_{\rm {P0}}$ and $L$,
which are by far the main contributors to the uncertainty of $C_3$. 
The error bars assigned for each thickness in Fig.~4 of the main text correspond 
to the variation of $C_3$ by changing $\Delta_{\rm {P0}}$ within $\pm 20 $ MHz range and $L$ within $\pm 2.5$ nm. 
For each thickness, the error bar associated to $C_{3}$ is the quadratic sum of the error due to $\Delta_{\textrm{P0}}$ and $L$.  

In order to give a final value for the extracted $C_3$, we make the final assumption that it does 
not vary with the cell thickness. This approximation supposes that retardation effects 
would not lead to an effective dependence of $C_3$ with $L$. The final value
of the cesium-sapphire $C_3$ coefficient is then assigned, together with its error bar, as an 
average weighted by the individual points of the error-bars. We find:
\begin{equation}
C_3=1.3\pm 0.1 \ \rm{kHz.}\mu \rm{m}^3
\end{equation}

\section{Theoretical calculation of $C_3^{\rm th}$}\label{Sec:Theory}

The level shift $V_{\rm a}(z)$ experienced by an atom in a state $|a\rangle$, placed in front of a 
dielectric surface with refractive index $n(\omega)$ is given by (see for example~\cite{Fichet1995}):
\begin{equation}\label{Eq:Va}
 V_{\rm a}(z) =
 - \frac{1}{4\pi\varepsilon_0} \sum_{\textrm{states }|b\rangle}
 \frac{n(\omega_{\rm ab})^2-1}{n(\omega_{\rm ab})^2+1}
 \frac{|\mu_{\rm x}^{\rm ab}|^2
 + |\mu_{\rm y}^{\rm ab}|^2
 +2|\mu_{\rm z}^{\rm ab}|^2}{16 z^3} \equiv - \frac{C_3^{\rm a}}{z^3}.
\end{equation}

The prefactor depending on the refractive index $n(\omega_{\rm ab})$ at the
frequency $\omega_{\rm ab}$ of an atomic transition
$|\rm a\rangle \rightarrow |\rm b\rangle$ accounts for the strength
of the image dipole formed inside the dielectric. It would be equal to 1 for
a perfectly reflective surface. The sum is over all states $|\rm b\rangle$
that are dipole-coupled to $|\rm a\rangle$,
and $\mu^{\rm ab}_{\rm x \ldots z}$ are the dipole matrix elements
corresponding to the dipole orientations along the $x$, $y$ and $z$ axis
respectively. Knowing the refractive index of the surface for a
range of frequencies and the dipole matrix elements of the corresponding 
transitions allows one to calculate the energy shift of a given state
due to van der Waals interaction, as given in Table~\ref{tab:schemes}.
In this table, the dipole matrix elements are theoretically calculated
values from Ref.\cite{Safronova2016} and Alkali Rydberg Calculator (ARC) 
Python package~\cite{Sibalic2016d}.  The refractive index of ordinary Sapphire
is calculated from the data in Ref.~\cite{Weber2003}. 
Using the values in Table~\ref{tab:schemes}, we finally obtain the theoretical van der Waals 
coefficient for the shift of  the  $6S_{1/2}\rightarrow 6P_{1/2}$ transition:
\begin{equation}
C_3^{\rm th}=C_3[6P_{1/2}]-C_3[6S_{1/2}]=1.10 \pm 0.03\textrm{ kHz}.\mu\textrm{m}^3.
\end{equation}

The expression~(\ref{Eq:Va}) used for $V_a(z)$ accounts only for the interaction of a dipole 
with its image in the sapphire. The interaction of dipole image with its dipole image, and higher, 
is neglected. The full potential would have the form~\cite{Hinds1991}:
\begin{eqnarray}
V^{\rm tot}_a(z)= -\frac{1}{4\pi \varepsilon_0}\sum_{\textrm{states }|b\rangle}&
&\left\{ \frac{|\mu_{\rm x}^{\rm ab}|^2+
|\mu_{\rm y}^{\rm ab}|^2
+2|\mu_{\rm z}^{\rm ab}|^2}{16} \sum_{m=0}^\infty \left[ \frac{1}{(z+mL)^3}+\frac{1}{(L-z+mL)^3}\right]\delta(\omega_{\rm ab})^{2m+1}\right. \nonumber\\
 &&\left. - \frac{|\mu_{\rm x}^{\rm ab}|^2+
|\mu_{\rm y}^{\rm ab}|^2
-2|\mu_{\rm z}^{\rm ab}|^2}{16} \sum_{m=1}^\infty \frac{2}{(mL)^3} \delta(\omega_{\rm ab})^{2m} \right\},
\end{eqnarray}
where $\delta(\omega_{\rm ab})= [n(\omega_{\rm ab})^2-1]/[n(\omega_{\rm ab})^2+1]$ 
is a factor depending on the material, and where the first
and second line correspond respectively to odd and even
reflections. 
Due to spherical symmetry 
$|\mu_{\rm x}|^2 = |\mu_{\rm x}|^2 = |\mu_{\rm z}|^2$
and the second term is zero. 
For the atoms in the middle of the cell, $z=L/2$ and 
the contribution of the $m=1$ term will  already be $1/3^3$ smaller
(closer to the surfaces, the suppression will be even stronger) due to the larger distance
between the atom and their image.
The extra factor $\delta^2\approx(0.5)^2$ leads to further suppression,
giving a total contribution on the order of $\sim$1\% compared to the
$m=0$ term taken into account in the potential $V(z)$ used in the main text.
Higher order reflections are even more suppressed. 
Thus the potential $V(z)$ used in the main text is a good approximation of the $V^{\rm tot}$
to within $\sim 3$\%.

\begin{table}[h!]
\caption{Dominant contributions to
the cesium-sapphire van der Waals constant $C_3$ for
$6S_{1/2}$ and $6P_{1/2}$ states, with listed  sources for 
dipole matrix elements $\mu^{\rm ab}$.}
	\label{tab:schemes}
	\centering
	\begin{tabular*}{0.5\linewidth}{@{\extracolsep{\fill}}ccccc}
		 \toprule
		 Hyperfine transition			& contribution to $C_3^{\rm{a}}$ (kHz.$\mu$m$^3$)	& Source \\
		 \midrule
		$6S_{1/2}\leftrightarrow 6P_{1/2}$	& 0.42 	& \cite{Safronova2016}	\\
		$6S_{1/2}\leftrightarrow 6P_{1/2}$	& 0.83	&	\cite{Safronova2016} \\
		$6S_{1/2}\leftrightarrow 7P_{3/2}$	& 0.01	& \cite{Safronova2016}	\\
		\midrule
		Total $C_3^{6S_{1/2}}=$ & 1.258(2)  \\
		\midrule
				$6P_{1/2}\leftrightarrow 6S_{1/2}$	& 0.42 & 	\cite{Safronova2016}	\\
		$6P_{1/2}\leftrightarrow 7S_{1/2}$	& 0.37	& \cite{Safronova2016}	\\
		$6P_{1/2}\leftrightarrow 8S_{1/2}$	& 0.02	&	\cite{Sibalic2016d} \\
		$6P_{1/2}\leftrightarrow 5D_{3/2}$	& 1.01	&	\cite{Safronova2016} \\
		$6P_{1/2}\leftrightarrow 6D_{3/2}$	& 0.37	&	\cite{Safronova2016} \\
		$6P_{1/2}\leftrightarrow 7D_{3/2}$	& 0.09 	& \cite{Safronova2016}	\\
		$6P_{1/2}\leftrightarrow 8D_{3/2}$	& 0.04	& \cite{Sibalic2016d}	\\
		$6P_{1/2}\leftrightarrow 9D_{3/2}$	& 0.02	&	\cite{Sibalic2016d} \\
		$6P_{1/2}\leftrightarrow 10D_{3/2}$	& 0.01	&	\cite{Sibalic2016d} \\
		\midrule
		Total $C_3^{6P_{1/2}}=$ & 2.36(3) & \\

		\bottomrule
	\end{tabular*}
\end{table}


%